\documentclass[aps,prb,reprint,a4paper,floatfix,showpacs,showkeys,superscriptaddress]{revtex4-1}

\usepackage{graphicx}
\usepackage{amsmath}
\usepackage{epstopdf}

\newcommand{\mytilde}{\raise.17ex\hbox{$\scriptstyle\mathtt{\sim}$}}

\begin{document}

\title{Charge Carrier Concentration Dependence of Encounter-Limited Bimolecular Recombination in Phase-Separated Organic Semiconductor Blends}

\author{Michael C. Heiber}
\email{heiber@mailaps.org}
\affiliation{Institut f{\"u}r Physik, Technische Universit{\"a}t Chemnitz, 09126 Chemnitz, Germany}
\affiliation{Center for Polymers and Organic Solids, University of California, Santa Barbara, California 93106, USA}

\author{Thuc-Quyen Nguyen}
\affiliation{Center for Polymers and Organic Solids, University of California, Santa Barbara, California 93106, USA}
\affiliation{Department of Chemistry and Biochemistry, University of California, Santa Barbara, California 93106, USA}

\author{Carsten Deibel}
\email{deibel@physik.tu-chemnitz.de}
\affiliation{Institut f{\"u}r Physik, Technische Universit{\"a}t Chemnitz, 09126 Chemnitz, Germany}

\date{\today}

\begin{abstract}
Understanding how the complex intermolecular- and nano-structure present in organic semiconductor donor-acceptor blends impacts charge carrier motion, interactions, and recombination behavior is a critical fundamental issue with a particularly major impact on organic photovoltaic applications. In this study, kinetic Monte Carlo (KMC) simulations are used to numerically quantify the complex bimolecular charge carrier recombination behavior in idealized phase-separated blends.  Recent KMC simulations have identified how the encounter-limited bimolecular recombination rate in these blends deviates from the often used Langevin model and have been used to construct the new power mean mobility model.  Here, we make a challenging but crucial expansion to this work by determining the charge carrier concentration dependence of the encounter-limited bimolecular recombination coefficient.  In doing so, we find that an accurate treatment of the long-range electrostatic interactions between charge carriers is critical, and we further argue that many previous KMC simulation studies have used a Coulomb cutoff radius that is too small, which causes a significant overestimation of the recombination rate.  To shed more light on this issue, we determine the minimum cutoff radius required to reach an accuracy of less than $\pm10\%$ as a function of the domain size and the charge carrier concentration and then use this knowledge to accurately quantify the charge carrier concentration dependence of the recombination rate.  Using these rigorous methods, we finally show that the parameters of the power mean mobility model are determined by a newly identified dimensionless ratio of the domain size to the average charge carrier separation distance. 
\end{abstract}



\maketitle

\section{Introduction}
Organic semiconductors have generated a lot of interest over the last two decades for their potential use in next generation light emitting diodes (OLEDs), photovoltaics (OPVs), photodetectors (OPDs), transistors (OFETs), and other electronic devices.  For photovoltaic and photodetector applications, the use of donor-acceptor blends has been shown to be particularly important for efficient operation.\cite{deibel2010review}  In such blends, two materials form a complex phase separated morphology, which is commonly called a bulk heterojunction (BHJ) structure.  The morphological details of the BHJ structure on the molecular-, nano-, and meso-scale have been shown to often have a major impact on device performance.\cite{vandewal2013,huang2014} Understanding in greater detail how these morphological features impact device operation is critical for continuing to improve the performance of these technologies.\cite{jackson2015}

One of the most important aspects of device operation that remains to be understood is the loss of charge carriers due to bimolecular recombination.  This loss occurs when an electron and a hole, which are not generated from the same absorbed photon, meet and recombine within the active later.  In simple systems, the result of this mechanism is a second-order reaction, in which the recombination rate is defined
\begin{equation}
	R_\text{rec} = -\frac{dn}{dt} = k_\text{br}np,
	\label{eqn:recomb}
\end{equation}
where $k_\text{br}$ is the bimolecular recombination rate coefficient, $n$ is the electron concentration, and $p$ is the hole concentration. This process is the dominant loss pathway in most well-performing OPVs and must be minimized in order to realize more efficient devices for renewable energy generation.\cite{cowan2012} Reducing the recombination rate increases the fill factor,\cite{mauer2010jpcl,credgington2012am} open-circuit voltage,\cite{shuttle2008prb,maurano2010,blakesley2011,credgington2012jpcl,collins2016} and ultimately the power conversion efficiency.  In addition, a low recombination rate depresses charge carrier mobility requirements\cite{stolterfoht2014} and allows devices to have a thicker active layer,\cite{clarke2012b,li2014wentao} which increases light absorption and makes it easier to manufacture  devices via roll-to-roll printing methods. 

To reduce the overall recombination rate, it is then imperative to understand the factors that control the magnitude of the rate coefficient ($k_\text{br}$).  In organic semiconductor films, the Langevin model is often used to describe the rate coefficient such that
\begin{equation}
	k_\text{br} \approx k_\text{L} = \frac{e}{\epsilon\epsilon_0} (\mu_e + \mu_h),
	\label{eqn:langevin}
\end{equation}
where $e$ is the elementary charge, $\epsilon$ is the dielectric constant, $\epsilon_0$ is the vacuum permittivity, and $\mu_e$ and $\mu_h$ are the electron and hole mobilities, respectively.\cite{langevin1903}  Both theoretical and experimental studies have shown that this model works well to describe the recombination rate in neat disordered organic semiconductors when the electric field is low.\cite{vanderholst2009,blom1997,pivrikas2005a,vanmensfoort2011,gorenflot2014}  However, in the more complicated BHJ structure, electrons are typically restricted to the acceptor phase and holes to the donor phase, and electron-hole recombination can only occur at a donor-acceptor interface. This spatial limitation on charge carrier motion and recombination location is expected to alter the recombination process compared to single phase films, in which electrons and holes are both present in the same material and can move freely and recombine anywhere in the film.  And experimentally, BHJ films have been found to show very complex recombination behavior that frequently deviates from the Langevin model.\cite{proctor2013b,lakhwani2014}  Most importantly, in some blends, the recombination rate has been found to be several orders of magnitude less than what is predicted by the Langevin model.\cite{pivrikas2005b,shuttle2008prb,deibel2008,scharber2010,clarke2011,murthy2013,albrecht2012,proctor2013a,wetzelaer2013,proctor2014a,roland2014}     
As a result, many of these studies have determined the Langevin reduction factor ($\zeta$), where
\begin{equation}
	\zeta = \frac{k_{br}}{k_\text{L}},
	\label{eqn:reduction_factor}
\end{equation}
to characterize the recombination behavior.  However, in other seemingly similar systems, recombination rates much closer to the Langevin model were observed.\cite{mozer2005c,dennler2006,scharber2010,clarke2011,clarke2012b,zalar2014}  Even blends with very similar morphology can have dramatically different recombination rates.\cite{clarke2011}

Understanding the origins of the reduced factor is critical for designing molecules and optimizing materials processing conditions to improve OPVs, but even after over 10 years of rigorous efforts, major fundamental questions still remain largely because experimentally disentangling the various contributing factors has been a significant challenge.  Accurately quantifying the recombination rate coefficient and the electron and hole mobilities under the same conditions itself is not trivial, but the largest difficulty is in controlling and limiting the number of independent variables between samples.  For example, in some BHJ films the the recombination rate was found to be proportional to the mobility of the slowest charge carrier,\cite{koster2006a,wetzelaer2013} and as a result the minimum mobility model was proposed,\cite{koster2006a} where
\begin{equation}
	k_\text{min} = \frac{e}{\epsilon\epsilon_0} \text{min}(\mu_e , \mu_h).
	\label{eqn:min_mobility}
\end{equation}
However, testing the generality of this model requires comparing the mobility dependence of many different blends with different electron and hole mobilities, and separating the effect of the electron and hole mobilities from the effect of morphological features is particularly challenging. Comparing different materials with different mobilities is often complicated by that fact that the blend morphology and other potentially important parameters may also be significantly different between samples.  The same challenge is present when trying to determine the impact of even simple morphological features such as the domain size.  Some experiments have observed that the recombination rate has a relatively weak domain size dependence,\cite{albrecht2012} but others have indicated a greater domain size dependence.\cite{roland2014}  However, it is difficult to control for changes to the electron and hole mobilities or other properties.

To help quantify the complex recombination behavior in these materials and develop better models, computational simulations and theoretical models have been an extremely valuable complement to experimental studies.  Among the available simulation and modeling tools, kinetic Monte Carlo (KMC) simulations have been a very useful tool for probing the behavior of systems that are too complex for analytical derivation.  This strategy has been most famously applied towards understanding charge transport in disordered semiconductors, through which the Gaussian disorder model was derived.\cite{bassler1993} In addition, a number of KMC simulation studies have yielded important results for understanding the charge generation and recombination processes in both neat and blend films.\cite{groves2013}  More specifically, \citeauthor{groves2008prb} used KMC simulations to show that the recombination rate in a simple phase separated system follows neither the Langevin model nor the minimum mobility model.\cite{groves2008prb} 

Recent KMC simulations were used to quantify the separate impact of the electron and hole mobilities and the domain size.\cite{heiber2015prl} Through this work, the new power mean model was constructed to describe the encounter-limited (diffusion-limited) recombination regime, 
\begin{equation}
\label{eq:power_mean}
k_\text{pm} =  \frac{e}{\epsilon\epsilon_0} f_1(d) 2 M_{g(d)}(\mu_e,\mu_h),
\end{equation}
where $f_1$ is a domain size dependent prefactor, $M_{g(d)}(\mu_e,\mu_h)$ is the power mean of the mobilities,
\begin{equation}
M_g(\mu_e,\mu_h) = \left(\frac{\mu_e^g+\mu_h^g}{2}\right)^{1/g},
\end{equation}
and $g$ is the domain size dependent power mean exponent. Physically, $f_1$ appears to be a morphological reduction prefactor that represents the reduction of the recombination rate due to the previously discussed spatial limitation on charge carrier recombination locations, and $g$ is a kind of weighting factor that determines how much the encounter probability depends on the magnitude of the lowest mobility.  Values for the power mean model parameters $f_1$ and $g$ were determined for domain sizes from 5 to 55 nm at a carrier concentration of $1 \times 10^{16}$ cm$^{-3}$.\cite{heiber2015prl} When the domain size was very small, $f_1$ and $g$ were both approximately equal to one and the resulting expression is equal to the Langevin model, but as the domain size increased both $f_1$ and $g$ decreased and deviations from the Langevin model were readily apparent.  However, these simulations have clearly shown that a greatly reduced recombination rate is not inherent to phase separated blends.  

To make further strides towards a general model for bimolecular recombination in phase separated blends, the carrier concentration dependence of the power mean model must also be determined.  However, accurate quantification of this dependence requires a carefully constructed and rigorous set of simulations.  One particularly important aspect is an accurate treatment of the long-range Coulomb interactions between the charge carriers.  Calculating these interactions is typically the most computationally intensive part of a KMC simulation,\cite{kimber2012,gagorik2014} and as a result, simplifications are often used.  Most commonly a cutoff radius is used, such that the interaction between any two charge carriers that are further apart than the cutoff radius is disregarded.  Early KMC studies used a fairly small cutoff radius of around 10~nm,\cite{watkins2005,yang2008,pershin2012} while others chose to disregard Coulomb interactions completely.\cite{nelson2003,lei2008,baidya2013}  Many later studies have used a larger cutoff radius in the range of 15-20~nm corresponding to the thermal capture radius, the distance at which the Coulomb potential becomes equal to $kT$.\cite{marsh2007,groves2008prb,meng2010,meng2011}  However, \citeauthor{casalegno2010} showed that using a cutoff radius at or below the thermal capture radius produces significantly underestimated device performance compared to a more accurate calculation using an Ewald sum method.\cite{casalegno2010}  To reduce this error, some studies have included interactions between all charge carriers in the lattice.\cite{deibel2009a,strobel2010,kimber2010,kimber2012,lyons2012,gagorik2014,jones2014}  However, this method introduces an implicit cutoff equal to half of the smallest lattice dimension, and if the lattice is not large enough, the same issue arises.

In this study, we determine the minimum Coulomb cutoff radius required to produce accurate results within 10\% error and then use this knowledge to precisely characterize the charge carrier concentration dependence of the encounter-limited recombination coefficient.  Based on these results, we show that the power mean model parameters, $f_1$ and $g$, are proportional to a newly identified dimensionless ratio of the domain size ($d$) to the average charge carrier separation distance ($d_s$), and with this finding, a more generalized form of the power mean model is constructed.  This updated model further highlights the complex effect that the domain size, charge carrier concentration, and charge carrier mobilities all have on the recombination rate, even in a simplified BHJ morphology. This improved fundamental understanding helps address a critical problem in organic photovoltaic applications but may also be broadly relevant for other systems and applications where nanoscale structure dictates reaction kinetics.

\section{Methods}
Expanding on previous methods,\cite{heiber2012jcp,heiber2015prl} KMC simulations were configured to simulate the recombination dynamics in the bulk of a BHJ film during a pump-probe experiment that does not include electrode contacts.  Using the Ising\_OPV v2.0 software tool,\cite{heiber2015b} BHJ morphologies were created with a 50:50 volumetric blend ratio on lattices with a final size of 200 by 200 by 200 or larger using an interaction energy ($J$) equal to 0.6~$kT$.  After phase separation, domain smoothing was applied with a smoothing threshold of 0.52, and the resulting morphologies had two bicontinuous pure phases with an equal donor and acceptor domain size.  The domain size of each morphology was characterized using the pair-pair correlation method.  More details about the morphology generation and characterization techniques can be found elsewhere.\cite{heiber2014prapp}  Seven morphology sets consisting of 96 independently generated morphologies were created with an average domain size ($d$) of approximately 5, 10, 15, 20, 30, 40, and 50~nm. 

The morphologies were then implemented into a three-dimensional lattice with a lattice constant of 1~nm.  Three-dimensional periodic boundary conditions were used, and no energetic disorder was included to simplify the calculations.  We have previously found that energetic disorder only affects the magnitude of the mobility without changing the fundamental mobility dependence of the recombination rate.\cite{heiber2015prl}
To start the pump-probe simulation, excitons were created uniformly throughout the lattice with a Gaussian excitation pulse having a pulse width of 100~ps and an intensity corresponding to an initial exciton concentration of $1 \times 10^{17}$~cm$^{-3}$ in the film.  Free charge carriers were created directly from excitons by placing electron-hole pairs across the nearest donor-acceptor interface with a separation distance of at least 30~nm.  For domain sizes ($d$) of 30~nm or greater, the initial separation distance was set to $d+5$~nm to maintain a homogeneous initial charge carrier distribution. 

Charge transport was simulated using the Miller-Abrahams (MA) model,\cite{miller1960} and Coulomb interactions were included between any two charges that are located within a specified Coulomb cutoff radius ($r_c$).  Electron hopping was restricted to acceptor sites and hole hopping was restricted to donor sites. Charge recombination was also implemented using the MA model with a large recombination prefactor of $10^{15}$~s$^{-1}$ to put the system in the encounter-limited regime.\cite{heiber2015prl}  To investigate the impact of the Coulomb cutoff radius, $r_c$ was varied from 20~nm up to half the lattice size.  With a cutoff radius less than or equal to half of the lattice size, possible periodic artifacts were avoided.  For morphologies with 50~nm domains, a larger lattice with 250~nm dimensions was needed to avoid morphological confinement effects\cite{heiber2014prapp} and allow larger cutoff radii to be tested.  For each simulation, 24 morphologies were randomly selected from the appropriate morphology set and run 4 times.  The hole concentration ($p$) was logged as a function of time, and with the lattice sizes used, the carrier concentration could be resolved over one and a half orders of magnitude, covering a range typical for steady state illumination intensities from 0.1 to 5 suns. The final hole concentration transients were then obtained by averaging the 96 individual transients for each parameter set.  

The numerical derivative of the hole concentration as a function of time ($t$) was used to calculate the simulated recombination coefficient:
\begin{equation}
k_\text{sim}(t) = -\frac{\frac{dp(t)}{dt}}{p(t)^2}.
\label{eqn:k_sim}
\end{equation}
In addition, the displacement of each carrier from its initial position was recorded over its lifetime, and the behavior of all electrons and holes was averaged to determine the average mean squared displacement for each carrier type.  The numerical derivative of the average mean squared displacement was then used to calculate the time-dependent diffusion coefficient for each carrier type.  Due to the three dimensional boundary conditions, the three-dimensional diffusion equation was used with the Einstein relation, and the average time-dependent zero-field mobility of each carrier type was determined,
\begin{equation}
\mu(t) = \frac{e}{6 k_\text{B} T} \frac{d\langle r(t)^2 \rangle}{dt}.
\label{eqn:mobility}
\end{equation}
This detailed analysis allows the determination of the relationship between the recombination rate coefficient and mobility of the electrons and holes at any time point along the transient and accounts for any changes in the mobility over time or at different carrier concentrations.  More information about the morphology generation process, KMC simulation parameters, and data analysis is provided in the Supplementary Information.\footnotemark[1]

\section{Results and Discussion}

\subsection{Effect of the Coulomb Cutoff Radius}
\begin{figure}[h]
\includegraphics[scale=1]{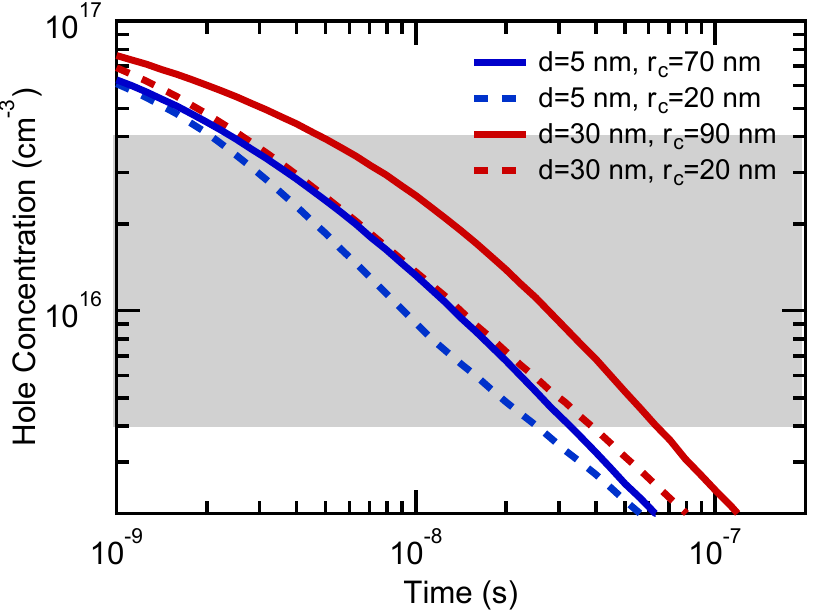}
\caption{\label{fig:transients}Hole concentration transients for domain sizes ($d$) of 5 nm (blue) and 30 nm (red) comparing the behavior obtained with a very large Coulomb cutoff radius (solid) and the thermal cutoff radius (dashed).  The highlighted gray area denotes the hole concentration range over which detailed analysis was performed.}
\end{figure}

First, we investigate how the Coulomb cutoff radius impacts the simulated recombination behavior under the conditions where the electron and hole hopping rates (mobilities) are equal.  Figure \ref{fig:transients} shows how reducing the Coulomb cutoff radius impacts the hole transients for small domain size (5~nm) and larger domain size (30~nm).  In both cases, when using a cutoff radius near the thermal capture radius (20~nm), the charge carriers recombine at a significantly faster rate than obtained with a large cutoff radius.  These results show that a cutoff radius of only 20~nm is not sufficient to obtain accurate recombination behavior.  Looking in more detail at the transients, it appears that the magnitude of the deviations differ depending on the domain size and the hole concentration.  To determine these differences quantitatively, detailed analysis was performed at several points over a range of hole concentrations from $4 \times 10^{15}$ to $4 \times 10^{16}$~cm$^{-3}$, which is highlighted by the gray box in Fig.~\ref{fig:transients}, for domain sizes from 5 to 50~nm.  This procedure was then repeated for different values of the Coulomb cutoff radius.

\begin{figure}[h]
\includegraphics[scale=1]{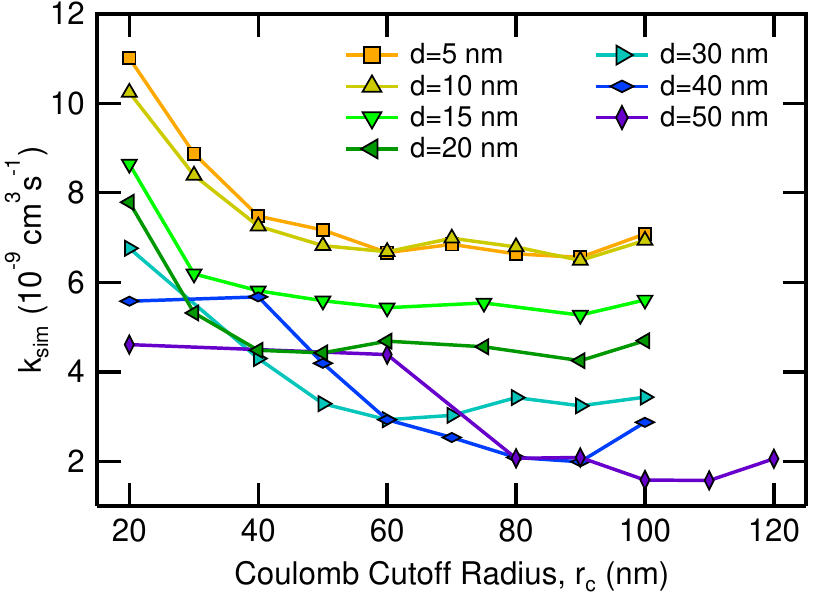}
\caption{\label{fig:k_analysis} Simulated recombination coefficients extracted from hole transients at a hole concentration of $2 \times 10^{16}$ cm$^{-3}$ for domain sizes ($d$) of 5, 10, 15, 20, 30, 40, and 50 nm as a function of the Coulomb cutoff radius.}
\end{figure}

Figure \ref{fig:k_analysis} shows how the recombination rate coefficient is affected by the choice of the Coulomb cutoff radius at $p=2 \times 10^{16}$ cm$^{-3}$.  With a cutoff radius of 20~nm, the charges disappear from the lattice significantly faster than with a much larger cutoff radius.  This shows that when the cutoff radius is too small, the simulation overestimates the recombination rate.  In some cases this overestimation can be as large as a factor of 2.  This result is consistent with previous simulation results by \citeauthor{casalegno2010}, which showed that using a small cutoff radius causes an underestimation of the photocurrent in an OPV device due to more recombination.\cite{casalegno2010}  However, it was not previously determined how large the cutoff radius must be to reach acceptable accuracy.  To do this, the results shown in Fig.~\ref{fig:k_analysis} were analyzed further.

As the Coulomb cutoff radius increases, the simplification artifact is reduced, and the recombination coefficient approaches the mathematically exact value.  However, since we do not know the exact value \textit{a priori}, the plateau value was used as a close approximation.  If the final three points were within 10\% of each other, the system was deemed to have reached the plateau, and the final two points were averaged to determine the plateau value. 
For each combination of domain size and charge carrier concentration, the minimum value of the Coulomb cutoff radius that produces a recombination coefficient within 10\% of the plateau value ($r_{c,10}$) was then determined.  As a result, the recombination coefficient obtained when using a Coulomb cutoff radius greater than or equal to $r_{c,10}$ should be within 10\% of the exact value.  This analysis was repeated for a number of different charge carrier concentrations from $p=4 \times 10^{15}$ to $4 \times 10^{16}$~cm$^{-3}$.

\begin{figure}[h]
\includegraphics[scale=1]{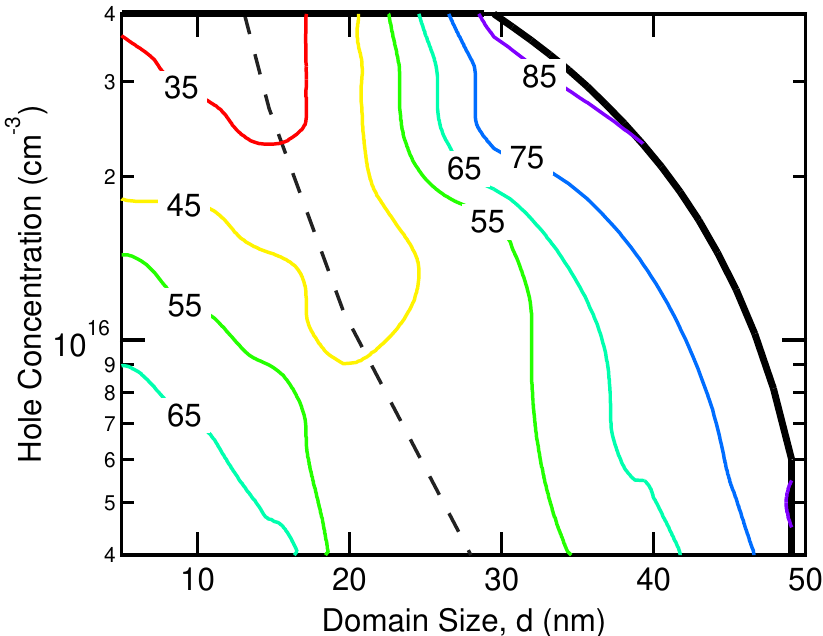}
\caption{\label{fig:cutoff_analysis} Contour plot showing the minimum Coulomb cutoff radius ($r_{c,10}$) extracted from recombination coefficient analysis as a function of domain size ($d$) and hole concentration ($p$).  The black dashed line shows the approximate transition between regime 1 and regime 2 ($d_s$=d).
}
\end{figure}

Figure \ref{fig:cutoff_analysis} shows a contour plot demonstrating interpolated values of $r_{c,10}$ as a function of the hole concentration and the domain size.  Based on the shape of the contour lines, two very different regimes are identified, one for small domain sizes and another for large domain sizes.  With small domains, the general trend is that as the carrier concentration decreases, the minimum cutoff radius increases.  This trend has also been identified by \citeauthor{vanderholst2009} when performing KMC recombination simulations for neat films.\cite{vanderholst2009}  In contrast, with larger domains, the minimum cutoff radius is mostly dependent on the domain size.  The estimated transition between the two regimes is shown by the black dashed line and is described below.  This data shows the conditions at which the small cutoff radius artifact is likely to have the greatest affect and provides guidelines for how future simulation studies can reduce this artifact.  In practice, the choice of the Coulomb cutoff radius has a dramatic impact on the calculation time, so the implementation of this approximation must be done carefully to make sure that the simulations are both tractable and produce valid conclusions.

\begin{figure}[h]
\includegraphics[scale=1]{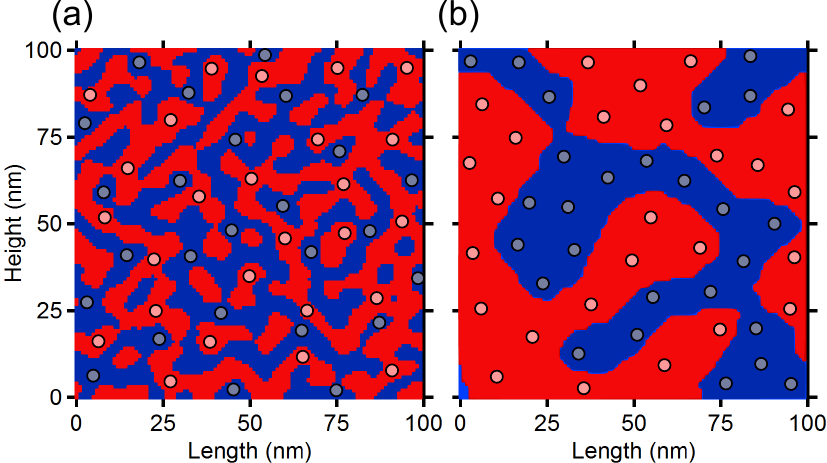}
\caption{\label{fig:regimes} Illustrations depicting (a) regime 1, in which the domains are smaller than the average separation distance between charge carriers, and (b) regime 2, in which the domains are larger than the average separation distance between charge carriers.}
\end{figure}

To understand the physical origins of the two regimes, we start by visualizing the state of the system under the two extreme conditions.  Figure \ref{fig:regimes} shows a 2-D representation of the two regimes.  In Fig.~\ref{fig:regimes}(a), the small domain size regime reveals a situation in which the domain size is smaller than the average separation distance between the charge carriers.  We estimate the average nearest neighbor distance of the charge carriers ($d_s$) assuming a random distribution of non-interacting particles,\cite{chandrasekhar1943}
\begin{equation}
d_s = \Gamma(4/3)\left[ \frac{3}{4 \pi (n+p)  } \right]^{1/3} = 0.55396 (n+p)^{-1/3}
\end{equation}
where $ \Gamma$ is the gamma function.  While this expression is not strictly valid because the electrons and holes are indeed interacting, it serves as a reasonable simple estimate.  In regime 1, the minimum cutoff radius is approximately proportional to the average separation distance.  This means that when the domains are small, the most important issue is whether or not the carriers can properly feel the interactions with their neighboring charge carriers.  As a result, when the carrier concentration decreases, the average separation distance increases, and the cutoff radius must be made correspondingly larger to include the interactions with the neighboring charges.  This requirement has also been identified in KMC recombination simulations on neat films.\cite{vanderholst2009} 

However, in regime 2 the domain size is larger than the average separation distance between carriers, as depicted in Fig.~\ref{fig:regimes}(b).  In this regime, the requirements for the minimum cutoff radius change dramatically.  If a small cutoff radius is used, a charge carrier in the interior of a domain will only feel the repulsive interactions of other nearby charge carriers with the same sign.  Only when the cutoff radius is significantly increased would this charge carrier also feel the attractive force of the opposite charge carriers across the domain interface.  

As the cutoff radius is increased even further, the charge eventually feels another smaller repulsive interaction from the next cluster of equal sign charge carriers. This behavior causes the recombination rate coefficient to show oscillatory behavior with increasing cutoff radius as can be observed in Fig.~\ref{fig:k_analysis} for 20, 30, 40, and 50~nm domains.  In all four cases, a minimum is observed when the cutoff radius is approximately equal to twice the domain size.  It is important to note that this minimum is not the true plateau as it may appear, especially for 50~nm domains.  The real plateau is not finally observed until the oscillations are dampened at a fairly large cutoff radius, and in some cases, the real plateau cannot be observed within our sampling space. In these cases (large domain size and large carrier concentration), the minimum cutoff radius cannot be determined. In addition, the magnitude of the oscillations increase with increasing carrier concentration, which explains the sharp transition between the two regimes at high carrier concentration and the less clear transition point at lower concentrations.  At low carrier concentrations, the oscillations can be less than the 10\% error threshold and have no visible impact on the minimum cutoff radius.  Nevertheless, the transition between the two regimes occurs approximately when the average charge separation distance equals the domain size, as shown by the black dashed line in Fig.~\ref{fig:cutoff_analysis}.  Understanding the physical origins of these two regimes highlights the very different electrostatic environments experienced by the charge carriers under different conditions, which may also alter the mobility and ultimately the recombination rate of the charge carriers in addition to dictating the minimum Coulomb cutoff radius.

\subsection{Charge Carrier Concentration Dependence of Recombination}

\begin{figure}[h]
\includegraphics[scale=1]{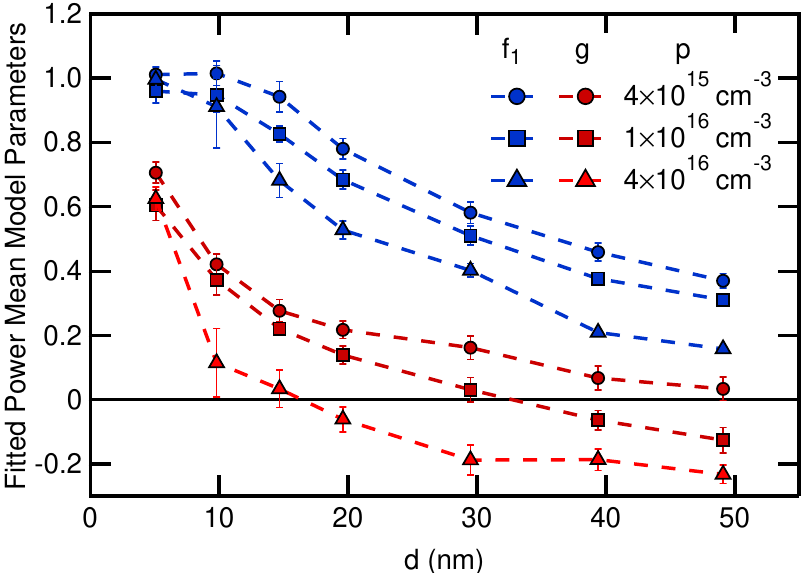}
\caption{\label{fig:fit_results1}Power mean model fitting results for parameters, $f_1$ and $g$, as a function of the domain size ($d$) at hole concentrations ($p$) of $4 \times 10^{15}$~cm$^{-3}$, $1 \times 10^{16}$~cm$^{-3}$, and $4 \times 10^{16}$~cm$^{-3}$.}
\end{figure}

Finally, to determine the charge carrier concentration dependence of the power mean mobility model, the relative magnitudes of the electron and hole mobilities were tuned by varying the hole hopping prefactor ($R_\text{0,h}$) from $10^{11}$ to $10^{15}$~s$^{-1}$ while holding the electron hopping prefactor ($R_\text{0,e}$) constant at $10^{13}$~s$^{-1}$.  With a sufficiently large Coulomb cutoff radius, the simulated recombination coefficient was determined as a function of both the domain size and the charge carrier concentration.  Under each set of conditions, the recombination coefficient was calculated at several points between $p=4 \times 10^{15}$ to $4 \times 10^{16}$~cm$^{-3}$.  All data was fit using Eqn.~\ref{eq:power_mean} to determine the fit parameters $f_1$ and $g$ (see Supplementary Information for exemplary data fits).\footnotemark[1]  Fig.~\ref{fig:fit_results1} shows how the fit parameters change as a function of domain size when analyzed at three different hole concentrations.  For all domain sizes, both $f_1$ and $g$ decrease at higher carrier concentrations.

\begin{figure}[h]
\includegraphics[scale=1]{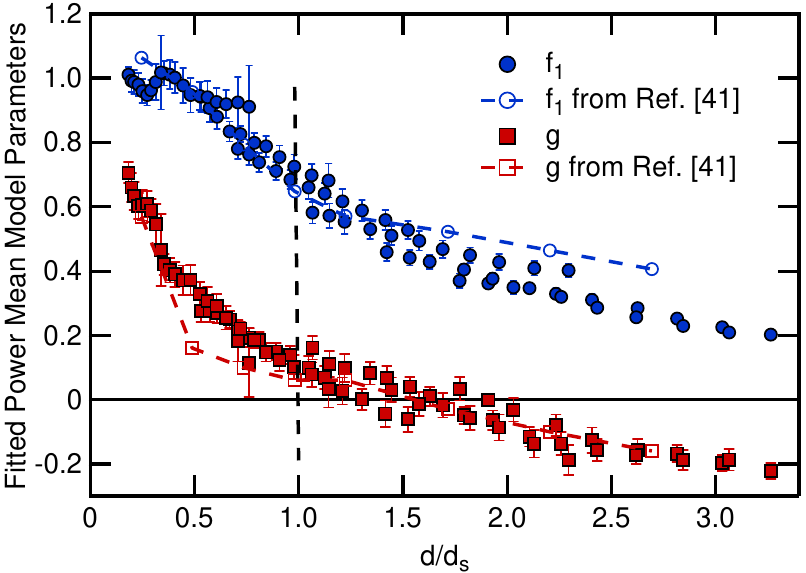}
\caption{\label{fig:fit_results2}Power mean model fitting results for parameters, $f_1$ and $g$, compared to previous data\cite{heiber2015prl} as a function of the dimensionless ratio of the domain size to the average charge carrier separation distance ($d/d_s$).  The black dashed line shows the approximate transition between regime 1 and regime 2.}
\end{figure}

Given the importance of the average charge carrier separation distance ($d_s$) in determining the two Coulomb interaction regimes above, the charge carrier concentration for each data set was translated into the corresponding value for $d_s$.  When plotting the fit parameters as a function of $d/d_s$, as shown in Fig.~\ref{fig:fit_results2}, we find that the recombination coefficient data obtained over a wide range of different domain sizes and carrier concentrations all come together to form one master curve.  From this analysis, we can conclude that both $f_1$ and $g$ are dependent on the dimensionless ratio between the domain size (d) and the average charge carrier separation distance ($d_s$).  In addition, the previously published data that was used to construct the power mean model also follows the same general trend but with minor deviations at some points likely due to insufficient Coulomb interactions.\cite{heiber2015prl}  Another interesting observation is that $f_1$ and $g$ have different behaviors in the two regimes identified in the Coulomb cutoff radius analysis.  In regime 1 where $d/d_s<1$, both $f_1$ and $g$ decrease quickly as $d/d_s$ increases, but in regime 2 where $d/d_s>1$,  both $f_1$ and $g$ show a much more gradual decrease.  The different trends observed in the two regimes might be due to the different electrostatic environments experienced by the charges in each regime.

As a result of this behavior, the power mean model can be updated to account for the effects of both the domain size and the charge carrier concentration.  The encounter-limited bimolecular recombination coefficient can now be defined,
\begin{equation}
k_\text{pm} =  \frac{e}{\epsilon\epsilon_0} f_1(d/d_s) 2 M_{g(d/d_s)}(\mu_e,\mu_h),
\label{eq:power_mean2}
\end{equation}
where both $f_1$ and $g$ are dependent on the dimensionless ratio,  $d/d_s$.  $f_1$ and $g$ may also depend on other parameters that are unknown at this time, but this more general form of the power mean model represents a significant step forward in understanding the bimolecular charge carrier recombination process in phase separated blends.  

In addition, in many BHJ blends used for OPVs, a simplified form of Eqn.~\ref{eq:power_mean2} can be used to describe the recombination rate under standard operating conditions.  When $d/d_s$ is between 1 and 2, $f_1 \approx 0.5$ and $g \approx 0$, and the power mean model can be simplified to
\begin{equation}
k_\text{e,OPV} =  \frac{e}{\epsilon\epsilon_0} \sqrt{\mu_e\mu_h},
\label{eq:geo_mean}
\end{equation}
With a typical domain size of \mytilde15~nm in optimized OPVs and a carrier concentration of $3 \times 10^{16}$~cm$^{-3}$, $d/d_s\approx1$, and this simplification is accurate to within 10-15\%.  However, if the domain size or carrier concentration is smaller than that, the simple geometric mean expression will begin to deviate significantly from the more complete power mean expression.  

Furthermore, even if real devices are not always truly in the encounter-limited recombination regime, the power mean model still defines the electron-hole encounter rate that will be part of the more complex rate equation.  For example, if we relax the assumption that electrons and holes always recombine immediately when they meet, as has also been posited in several previous studies,\cite{koster2005,groves2008prb,hilczer2010,burke2015} and define the electron-hole pairs that do form to have a finite electron-hole recombination rate ($k_\text{rec}$) and a dissociation rate ($k_\text{diss}$), the overall bimolecular recombination rate coefficient can be defined,\cite{ferguson2011}
\begin{equation}
k_\text{br} =  \frac{k_\text{rec}}{k_\text{diss}}k_\text{pm}.
\label{eq:k_br}
\end{equation}
In such a case, the power mean model, which describes the electron-hole encounter rate, is still a factor of general importance for describing the bimolecular recombination rate in BHJ blends over a very wide range of conditions.  In addition, the measured reduction factor ($\zeta$) would then be determined by the ratio of the electron-hole recombination rate coefficient to the dissociation rate coefficient multiplied by the small inherent reduction due to the phase separated morphology that is captured in the power mean model.

\section{Conclusions}
To conclude, rigorous KMC simulations were performed on model phase separated morphologies to identify the charge carrier concentration dependence of the encounter-limited bimolecular recombination rate.  In doing so, we find that correctly accounting for the long-range Coulomb interactions is critical and that using the small cutoff radius suggested by many previous studies results in a significant overestimation of the recombination rate.  Exploring this issue further, we determine the minimum cutoff radius required to reach an accuracy of less than $\pm10\%$ as a function of the domain size and the carrier concentration and identify two distinct regimes that reveal the presence of dramatically different electrostatic environments.  Then, using a sufficiently large cutoff radius, we find that the parameters of the power mean model are determined by the dimensionless ratio of the domain size to the average charge carrier separation distance, which results in a new more generalized form of the power mean model.

This work emphasizes the importance of correctly accounting for the long-range Coulomb interactions despite the often significant computational cost and provides clear guidelines for future KMC recombination simulation studies.  But most importantly, the newly constructed power mean model represents a major step forward in constructing a general model for the bimolecular recombination rate in organic semiconductor blends.  Understanding the fundamental structure-property relationships in these materials is critical for correctly interpreting the experimental observations in these complex systems and for future computational materials design.

M.C.H. and C.D. acknowledge funding by Deutsche Forschungsgemeinschaft (DFG) grant DE830/13-1. M.C.H and T.-Q.N. acknowledge funding by the Office of Naval Research (ONR) grant \#N000141410076.  We also thank Dr. Philipp Cain for assistance with computational resources and Niva A. Ran for insightful discussion and feedback on the manuscript.

\footnotetext[1]{See Supplemental Material for morphology generation details, a full list of simulation parameters, and supplementary results.}



%

\end{document}


\title{Supplementary Information for Charge Carrier Concentration Dependence of Encounter-Limited Bimolecular Recombination in Phase-Separated Organic Semiconductor Blends}

\author{Michael C. Heiber}
\email{heiber@mailaps.org}
\affiliation{Institut f{\"u}r Physik, Technische Universit{\"a}t Chemnitz, 09126 Chemnitz, Germany}
\affiliation{Center for Polymers and Organic Solids, University of California, Santa Barbara, California 93106, USA}

\author{Thuc-Quyen Nguyen}
\affiliation{Center for Polymers and Organic Solids, University of California, Santa Barbara, California 93106, USA}
\affiliation{Department of Chemistry and Biochemistry, University of California, Santa Barbara, California 93106, USA}

\author{Carsten Deibel}
\email{deibel@physik.tu-chemnitz.de}
\affiliation{Institut f{\"u}r Physik, Technische Universit{\"a}t Chemnitz, 09126 Chemnitz, Germany}

\date{\today}

\begin{abstract}
\end{abstract}

\pacs{}


\maketitle


\section{KMC Simulation Details}

The KMC simulation methodology used in this study was explained in more detail in previous papers,\cite{heiber2012jcp,heiber2015prl} but a summary of the most important aspects are provided here. The model morphologies were used to define the donor and acceptor sites on a three-dimensional lattice with a lattice constant ($a$) of 1~nm and periodic boundary conditions were used in all three directions. To start the simulation, excitons were created with uniform probability throughout the lattice with a Gaussian excitation pulse having a pulse width of 100~ps and an intensity corresponding to an initial exciton concentration of $1 \times 10^{17}$~cm$^{-3}$. 

Exciton diffusion was implemented using the F\"orster resonance energy transfer model,
\begin{equation}
R_{ij,\text{exh}} = R_{0,\text{exh}}\left(\frac{a}{d_{ij}}\right)^6 f_\text{B}(\Delta E_{ij,\text{exh}}),
\end{equation}
where $R_{0,\text{exh}}$ is the exciton hopping prefactor, $d_{ij}$ is the distance between sites, 
\begin{equation}
\label{eq:boltzmann}
f_\text{B}(\Delta E_{ij}) = \begin{cases}  \exp{\left(\frac{-\Delta E_{ij}}{kT} \right)} & \Delta E_{ij}> 0 \\
1 & \Delta E_{ij}\le 0 \end{cases},
\end{equation}
and $\Delta E_{ij,\text{exh}}$ is the change in potential energy for exciton hopping,
\begin{equation}
\Delta E_{ij,\text{exh}} = E_{j,\text{singlet}}-E_{i,\text{singlet}}.
\end{equation}
Exciton hopping events were calculated to sites up to 4~nm away from the starting site. In addition, the exciton relaxation time defines the lifetime of the excited state and is used to calculate the exciton relaxation rate,
\begin{equation}
R_\text{exr} = 1/\tau_\text{ex},
\end{equation}
where $\tau_{ex}$ is the exciton lifetime.

The complexities of charge separation were bypassed to create free charge carriers directly from excitons. To do this, electron-hole pairs were created across the interface with a separation distance of 30~nm by restricting exciton creation to within 30~nm of an interface and executing an ultrafast long-range charge transfer event.  In the simulations on morphologies with 40 and 50~nm domains, this separation distance was increased to 45 and 55~nm, respectively, to ensure a homogeneous initial charge carrier distribution.  This is important because when the initial separation distance is smaller than the domain size, the initial carrier concentration is higher closer to the interface, which results in a change in the recombination behavior at short times (high carrier concentrations).  Long range charge transfer (exciton dissociation) was implemented using the simplified Miller-Abrahams model where charge transfer is always energetically favorable,
\begin{equation}
R_{ij,\text{exd}} = R_{0,\text{exd}} \exp{(-2 \gamma_\text{ex} d_{ij})}
\end{equation}
where $R_{0,\text{exd}}$ is the exciton dissociation prefactor and $\gamma_{ex}$ is the inverse exciton localization parameter.  Exciton dissociation events were only calculated for pair of sites with a separation distance that is within 1~nm of the target charge separation distance.

Charge motion was simulated using the Miller-Abrahams model.  For electrons, 
\begin{equation}
R_\text{ij,\text{elh}} = R_\text{0,\text{e}} \exp({-2 \gamma_\text{ch} d_{ij}}) f_\text{B}(\Delta E_{ij,\text{elh}})
\end{equation}
where $R_\text{0,\text{e}}$ is the electron hopping prefactor, $\gamma_\text{ch}$ is the charge localization parameter, and
\begin{equation}
\Delta E_{ij,\text{elh}} = E_{i,\text{LUMO}}-E_{j,\text{LUMO}}+ \Delta E_{C,ij}-Fd_{ij},
\end{equation}
where $E_{i,\text{LUMO}}$ and $E_{j,\text{LUMO}}$ are the initial and final site energies drawn from the density of states distribution, $\Delta E_{C,ij}$ is the change in Coulomb potential that would occur for hopping from site $i$ to site $j$, and $F$ is the electric field. Analogous expressions are used to calculate the hole hopping rate. Coulomb interactions were included between charges within a cutoff radius ($r_c$).  The change in Coulomb potential is calculated,
\begin{equation} \Delta E_{C,ij} = E_{C,j}-E_{C,i}, \end{equation}
where
\begin{equation} E_{C,i} = \sum_{k=1, k\neq i}^N \frac {q_i q_k}{4 \pi \epsilon \epsilon_0 d_{ik}} \qquad d_{ik} \leq r_c, \end{equation}
given $N$ nearby electrons and holes.  An analogous expression is used to calculate the Coulomb potential for the final state by assuming that the charge of interest is positioned on site j. Electron hopping was restricted to acceptor domains and hole hopping was restricted to donor domains. Charge hopping events were calculated for sites up to 3~nm away from the starting site. 

When an electron and a hole come close together, the charge recombination event is enabled. Charge recombination was also implemented using the Miller-Abrahams model similar to charge hopping,
\begin{equation}
R_{ij,\text{rec}} = R_{0,\text{rec}} \exp{(-2 \gamma_{ch} d_{ij})},
\end{equation}
where $R_{0,rec}$ is the recombination prefactor, which was held constant at a large value of $10^{15}$~s$^{-1}$ to ensure that recombination dominates over re-dissociation.  Charge recombination events were also calculated for sites up to 3~nm away from the starting site.  In addition, the selective recalculation method \cite{heiber2012jcp} was used with a recalculation cutoff radius of 5~nm. A full list of parameters is provided in Table~\ref{tab:KMC_params}.

\begin{table}[h]
\begin{ruledtabular}
\caption{KMC Simulation Parameters}
\label{tab:KMC_params}
\begin{tabular}{ll}
Lattice constant, $a$ & 1 nm \\
Temperature, $T$ & 300 K \\
Dielectric constant, $\epsilon$ & 3.5 \\
Exciton lifetime, $\tau_{ex}$ & 500 ps \\
Exciton hopping prefactor, $R_{0,\text{exh}}$ & $10^{12}$ s$^{-1}$ \\
Exciton localization, $\gamma_\text{ex}$ & 0.1 nm$^{-1}$ \\
Exciton dissociation prefactor, $R_{0,\text{exd}}$ & 10$^{16}$ s$^{-1}$ \\
Electron hopping prefactor, $R_{0,\text{e}}$ &  10$^{13}$ s$^{-1}$ \\
Hole hopping prefactor, $R_{0,\text{h}}$ & variable \\
Charge localization, $\gamma_\text{ch}$ & 2 nm$^{-1}$ \\
Charge recombination prefactor, $R_{0,\text{rec}}$ & 10$^{15}$ s$^{-1}$ \\
Electric field, $F$ & 0 Vm$^{-1}$\\
Coulomb cutoff radius, $r_\text{c}$ & variable \\
\end{tabular}
\end{ruledtabular}
\end{table}

\section{Morphology Details}
Seven morphology sets (MS1, MS2, MS3, MS4, MS5, MS6, MS7) were generated with different domain sizes of approximately 5, 10, 15, 20, 30, 40, and 50~nm using the Ising\_OPV v2.0 software tool.\cite{heiber2015b} More detailed information on the input parameters and characterization of each morphology set is provided in Table~\ref{tab:morphology} and more details about the morphology generation and analysis methods are published elsewhere.\cite{heiber2014prapp}

\section{Supplementary Results}
\begin{figure*}[ht]
\includegraphics[scale=1]{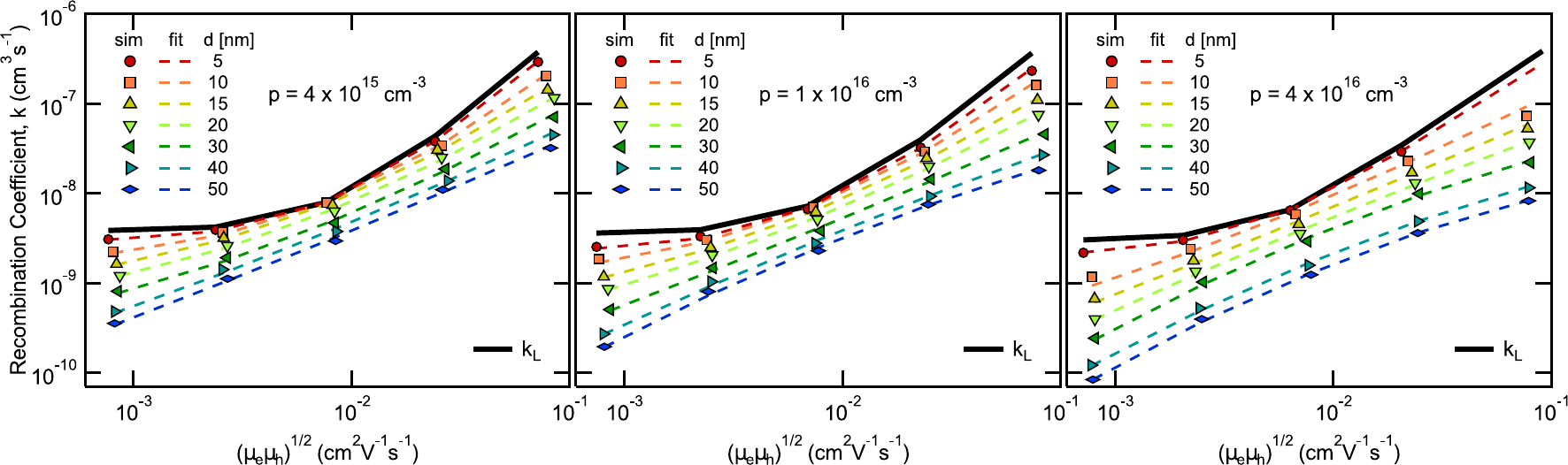}
\caption{\label{fig:SI_fits}Fitted recombination coefficient data from recombination simulations with varying hole mobility and domain size with transient analysis at $p = 4 \times 10^{15}$~cm$^{-3}$, $p = 1 \times 10^{16}$~cm$^{-3}$, and $p = 4 \times 10^{16}$~cm$^{-3}$ compared to the calculated Langevin recombination coefficient (solid black line).}
\end{figure*}

All simulated recombination coefficient data was fit with the equation,
\begin{equation}
k_\text{sim} =  \frac{e}{\epsilon\epsilon_0} f_1 2 M_{g}(\mu_e,\mu_h).
\label{eqn:fit}
\end{equation}
where $f_1$ is a prefactor and $M_{g}(\mu_e,\mu_h)$ is the power mean (generalized mean),
\begin{equation}
M_g(\mu_e,\mu_h) = \left(\frac{\mu_e^g+\mu_h^g}{2}\right)^{1/g},
\end{equation}
with a power mean exponent $g$.
Fig.\ \ref{fig:SI_fits} shows the resulting fits for $p = 4 \times 10^{15}$~cm$^{-3}$, $p = 1 \times 10^{16}$~cm$^{-3}$, and $p = 4 \times 10^{16}$~cm$^{-3}$.  The resulting fitted parameters $f_1$ and $g$ and their uncertainties are shown in the main article.  These results show that the power mean in Eqn.\ \ref{eqn:fit} is a very accurate description of the recombination behavior.

\begin{table*}[ht]
\caption{Morphology Set Information}
\label{tab:morphology}
\begin{tabular}{ l | c | c | c | c | c  | c | c | c}
\hline \hline
& MS1 & MS2 & MS3 & MS4 & MS5 & MS6 & MS7 \\
\hline
Initial lattice dimensions & 200 & 100 & 67 & 50 & 34 & 25 & 25 \\
Monte Carlo steps & 380 & 380 & 380 & 380 & 380 & 380 & 380 \\
Rescale factor & N/A & 2 & 3 & 4 & 6 & 8 & 10 \\
Smoothing threshold & 0.52 & 0.52 & 0.52 & 0.52 & 0.52 & 0.52 & 0.52 \\
\hline
Final lattice dimensions & 200 & 200 & 201 & 200 & 204 & 200 & 250 \\
Domain size, $d$ & 5.14$\pm$0.02 & 9.82$\pm$0.03 & 14.7$\pm$0.1 & 19.6$\pm$0.2 & 29.4$\pm$0.6 & 39$\pm$1 & 49$\pm$1 \\
Interfacial area/volume & 0.343$\pm$0.0004 & 0.183$\pm$0.0005 & 0.124$\pm$0.0008 & 0.093$\pm$0.0009 & 0.062$\pm$0.001 & 0.046$\pm$0.001 & 0.037$\pm$0.001 \\
Tortuosity & 1.10$\pm$0.01 & 1.10$\pm$0.02 & 1.10$\pm$0.02 & 1.10$\pm$0.03 & 1.11$\pm$0.04 & 1.11$\pm$0.05 & 1.11$\pm$0.06 \\
\hline \hline
\end{tabular}
\end{table*}

%